\documentclass[twocolumn,showpacs,preprintnumbers,amsmath,amssymb,ssc]{revtex4}
\usepackage[dvips]{color,graphics,epsfig,rotating}
\usepackage{graphicx}
\usepackage{dcolumn}
\usepackage{bm}

\begin{document}

\title{Ab Initio Phonon Dispersions for PbTe}

\author{Jiming An}
\affiliation{Materials Science and Technology Division,
Oak Ridge National Laboratory, Oak Ridge, Tennessee 37831-6114} 
\affiliation{Wuhan University of Technology, Wuhan, China}

\author{Alaska Subedi}
\affiliation{ Department of Physics and Astronomy, University of
Tennessee, Knoxville, Tennessee 37996-1200, USA}
\affiliation{Materials Science and Technology Division,
Oak Ridge National Laboratory, Oak Ridge, Tennessee 37831-6114} 

\author{D.J. Singh}
\affiliation{Materials Science and Technology Division,
Oak Ridge National Laboratory, Oak Ridge, Tennessee 37831-6114} 

\date{\today} 

\begin{abstract}
We report first principles calculations of the phonon dispersions
of PbTe both for its observed structure and under compression.
At the experimental lattice parameter we find a near instability of the
optic branch at the zone center, in accord with experimental observations.
This hardens quickly towards the zone boundary. There is also a very
strong volume dependence of this mode, which is rapidly driven away
from an instability by compression. These results are discussed in
relation to the thermal conductivity of the material.
\end{abstract}

\pacs{63.20.dk,63.20.Ry,64.70.kg}

\maketitle

\section{introduction}

The thermoelectric performance of a material is quantified using
a dimensionless, temperature dependent figure of merit, 
$ZT=\sigma S^2 T / \kappa$,
where $\sigma$ is the electrical conductivity, $\kappa$ is the
thermal conductivity, and $S$ is the thermopower.
Thus it is clear that high thermoelectric performance is generally found in
materials with low thermal conductivity,
and in fact various strategies for lowering thermal conductivity
based on phonon scattering
have been employed. \cite{ioffe,rowe,slack,cahill}
These include alloying to produce mass disorder as in SiGe,
the use of materials with large complex unit cells,
and the use of materials with rattling ions, such as filled clathrate
materials. \cite{cahill}
In this regard rocksalt
structure PbTe is a remarkable thermoelectric material. It is
based on a small, high symmetry unit cell, has high coordination of
both atomic species (6-fold), and contains no obvious rattling ion.
Nonetheless, $n$-type PbTe has high values of $ZT$ above room temperature
and as such was used in spacecraft power generators.
\cite{wood,gelbstein}
It also has low thermal conductivity, reported as 2.3 W/(m K) at
ambient temperature. \cite{joffe}
More recently, a family of so-called LAST compounds based on intergrowths
of PbTe and Ag$_x$SbTe$_2$ have been shown to have values of $ZT$
well above unity, \cite{hsu}
and additionally quantum dot systems based on PbTe
have also shown high values of $ZT$. \cite{harman}

Thermal properties of materials in general rest on the phonon dispersions.
While electronic and related properties of
PbTe has been rather extensively studied using first principles
methods, \cite{conklin,rabe1,waghmare,bilc,hoang}
the phonon dispersions have not been reported. The purpose
of this paper is to report these dispersions, as obtained within
density functional theory.

\section{approach}

The calculations reported here were done within the framework of
density functional theory primarily using the local density approximation.
The quantum espresso
code was used with norm conserving pseudopotentials to obtain
the phonon dispersions via linear response. \cite{qe}
Convergence tests were performed to determine the needed planewave
basis cut-off and zone sampling. The results shown used
a planewave basis set with a 60 Ry cutoff and an 6x6x6
special {\bf k}-points Brillouin zone sampling.
We performed convergence tests with a 75 Ry cutoff and an 8x8x8
special {\bf k}-points Brillouin zone sampling but found no significant
differences in phonon frequencies.
In addition, some calculations were done with the linearized augmented
planewave (LAPW) method including local orbitals.
\cite{singh-book}
These provide a check for the pseudopotential results.
Calculations were performed as a function of lattice parameter.
The calculated LAPW LDA lattice parameter is 6.39\AA, which is slightly
more than 1\% smaller than
the experimental room temperature lattice parameter of 6.464\AA,
\cite{dalven}
and within 1\% of the 120K value of 6.438\AA. \cite{noda}
For comparison the calculated LAPW lattice parameter within the generalized
gradient approximation of Perdew, Burke and Ernzerhof \cite{pbe}
is 6.57\AA,
which as is sometimes that case
for materials with heavy elements \cite{filippi}
is further from experiment than the LDA.
We therefore use the LDA, and perform calculations as a function
of lattice parameter. Assuming that PbTe behaves like other
materials with soft phonons, comparison with experiment should
be done for the phonon dispersions calculated at the experimental
lattice parameter. \cite{singh-92}

\section{results and discussion}

\begin{figure}[tb]
\includegraphics[width=3.3in,angle=0]{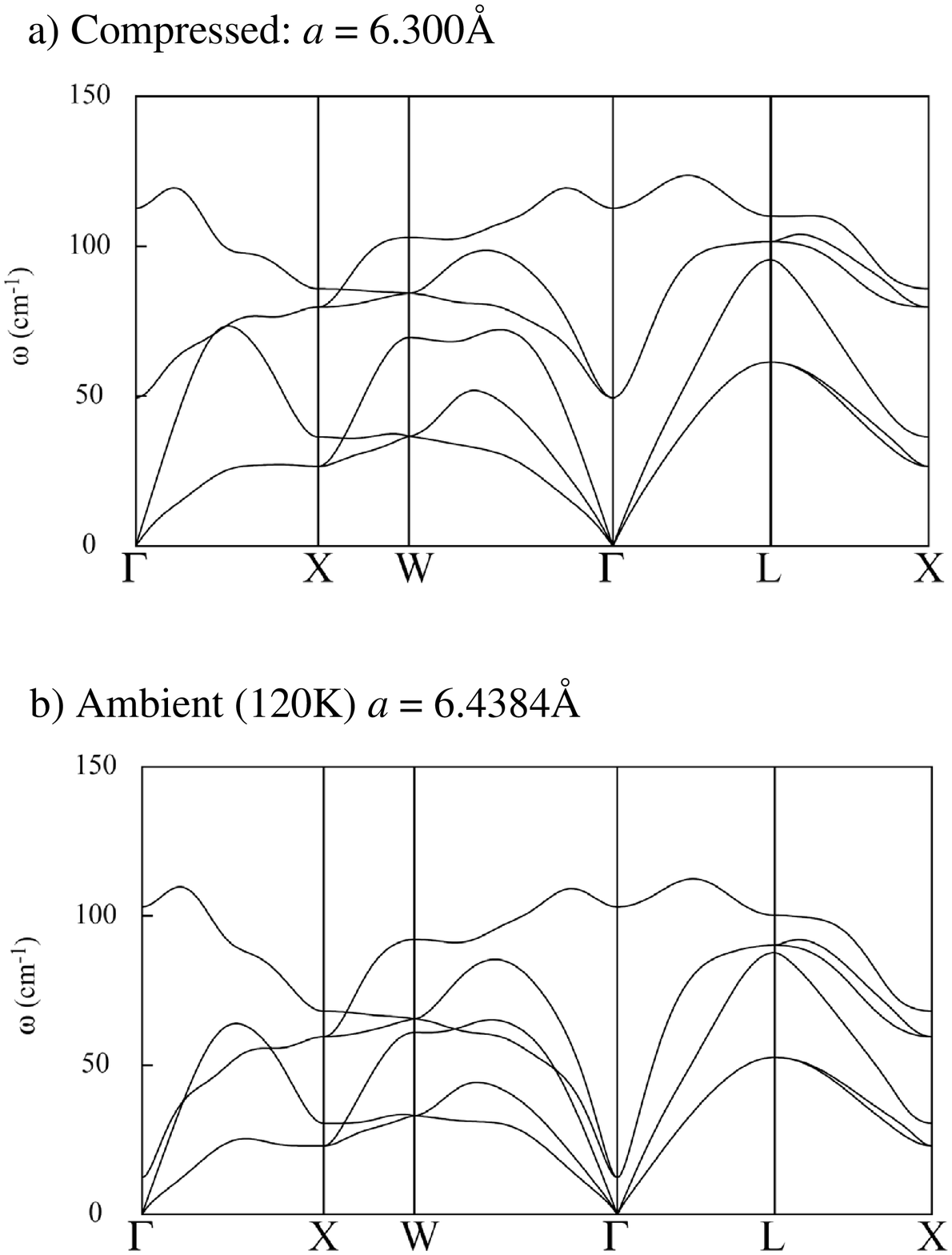}
\caption{\label{fig-phonon} 
Calculated phonon dispersions of PbTe at a lattice parameter of
$a$=6.30\AA~(top) and at the experimental 120K lattice parameter,
$a$=6.4384\AA~(bottom).
}
\end{figure}

Our main results are the phonon dispersions, which are given
in Fig. \ref{fig-phonon}.
As may be seen, PbTe is near a zone center phonon instability,
associated with the transverse optic (TO) mode. This is an inversion
center breaking phonon, and would lead to a ferroelectric
ground state if it were actually unstable.
This is not a surprising result and in particular
is in accord with experimental studies of NaCl structure group IV
tellurides. The related materials SnTe and GeTe are in fact
ferroelectric, with rhombohedral ground states corresponding to freezing
in of the soft TO mode with displacements along a [111] direction.
\cite{iizumi,pawley,lucovsky}

The ferroelectricity of GeTe has been studied in detail using first
principles calculations by Rabe and co-workers, \cite{rabe2,rabe3,rabe4}
who showed quantitative agreement between the predictions of LDA calculations
and experiment.
Bussmann-Holder used model calculations to investigate the phase
diagrams of various alloys including PbTe-SnTe and PbTe-GeTe
and made arguements for the nearness of PbTe to ferroelectricity.
\cite{bussmann}
In fact it was known that PbTe shows a dielectric constant that
increases at low $T$ as well as a TO phonon mode that decreases
in frequency as the temperature is lowered,
\cite{pawley,burkhard}
although the detailed $T$ dependence is complicated by a doping
level dependence.
In any case, Burhard and co-workers \cite{burkhard}
obtained obtained a very soft frequency of 17.5 cm$^{-1}$
from optical measurements on PbTe films. This value is consistent
with the very low value of 12 cm$^{-1}$ that we obtain at the
reported experimental lattice parameter. This difference of
$\sim$ 5 cm$^{-1}$ is well within the usual errors of LDA calculations.

\begin{figure}[tb]
\includegraphics[height=3.2in,angle=270]{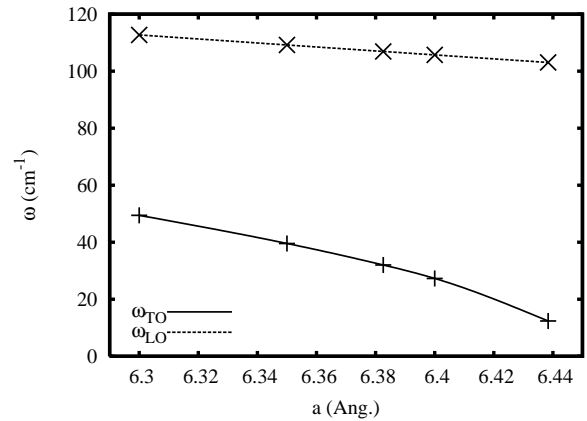}
\caption{\label{fig-vol} 
Zone center transverse and longitudinal
optical phonon frequency of PbTe as a function of lattice parameter.
}
\end{figure}

\begin{figure}[tb]
\includegraphics[width=3.2in,angle=0]{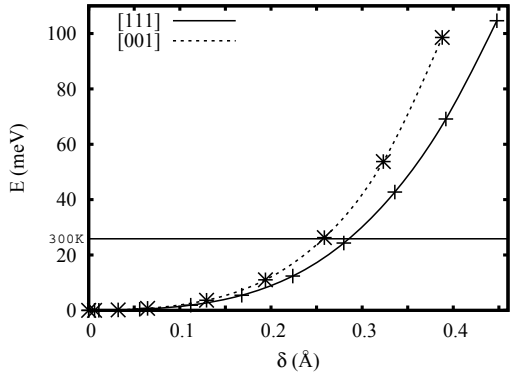}
\caption{\label{fig-en} 
Calculated energy as a function of phonon amplitude for the $\Gamma$
point TO mode, with displacements along the tetragonal and rhombohedral
directions for the experimental 300K lattice parameter of 6.464\AA.
The horizontal line corresponds to $kT$ for $T$=300K.
}
\end{figure}

Comparing the two panels of Fig. \ref{fig-phonon}, one may note
a strong volume dependence of the TO phonon frequency around the zone center.
This frequency at $\Gamma$
is plotted as a function of lattice parameter in Fig. \ref{fig-vol}.
As may be seen there is a very strong volume dependence to the soft TO mode.
Compressing the lattice parameter to 6.40 \AA, which is less than 0.6\%
smaller than the 120K value, more than doubles the TO frequency from
12 cm$^{-1}$ to 27 cm$^{-1}$.
The LO mode on the other hand shows a quite ordinary, modest volume
dependence.
While such strong volume dependences of soft
mode phonon frequencies are unusual
they have been noted before in complex oxide ferroelectrics,
such as Pb(Zr,Ti)O$_3$, BaTiO$_3$ and KNbO$_3$. \cite{samara,singh-92,fornari}
The present results show that there is a similar behavior in PbTe
in spite of the much simpler structure of this material in comparison
with the oxide ferroelectrics.

Returning to the issue of the thermoelectric properties of PbTe,
the material clearly does not involve alloying with different
mass ions or a large complex unit cell to lower thermal conductivity.
Furthermore, the acoustic modes are not anomalously soft. 
One mechanism that has been discussed in the context of thermoelectric
materials is the use of so-called rattlers. These are loosely bound
atoms in a crystal structure that couple to the low frequency part of the
phonon spectrum leading to scattering.
One may ask then if there is an analogy between the soft TO vibrations
near the zone center and a material containing rattlers.
One such connection
can be made based on the strong volume dependence. Longitudinal
acoustic (LA)
vibrations are compressive waves and so the strong volume dependence
of the TO mode implies strong
anharmonic coupling between the TO branch and the
LA phonons.
Furthermore, we find that the TO modes are highly anharmonic themselves.
Fig. \ref{fig-en} shows the calculated energy as a function of TO
mode amplitude at the experimental volume.
As may be seen the energy surface remains very soft up to large amplitude
$\sim$ 0.1 \AA, and then starts to stiffen in a quite anharmonic way.
This may be seen, {\em e.g.} from the difference between [001] and [111]
energies, which arises due to the anharmonicity. For large amplitudes
the softest direction is [111] in agreement with the experimental
ground state of GeTe.
However, even for [001] displacements the energy surface is very
soft implying large amplitude vibrations even at room temperature.
However, it should be noted that the TO mode is only soft near the
zone center, and so the mean square displacement will be less than
that implied by the position of the $kT$=300K line on Fig. \ref{fig-en}.
This also means that unlike rattling modes in clathrates, for example,
the soft TO modes in PbTe are strongly ${\bf k}$-dependent, similar to the 
acoustic modes.

\section{conclusion}

While the above results
do not provide a quantitative description of the thermal
conductivity of PbTe they do show that there
are unusual features of the phonon spectrum
that are generally associated with reduced thermal conductivity. In
particular we find very soft TO phonon branches near the zone center
as may be expected in a material near ferroelectricity. Importantly,
these modes are very volume dependent which implies strong anharmonic
coupling to longitudinal acoustic modes.
Furthermore, although the TO modes are quite {\bf k}-dependent,
which is not an indicator of low thermal conductivity, they are
also highly anharmonic, as seen from the Fig. \ref{fig-en}. This normally
will lead to increased scattering at high temperature and accordingly 
reduced thermal conductivity.

\acknowledgements

We are grateful for helpful discussions with Jihui Yang (General Motors)
and M.H. Du.
This work was supported by the Department of Energy, EERE,
Propulsion Materials Program and the ORNL LDRD Program and
by the Office of Naval Research (UT).

\end{document}